# Different Rise Times of Atomic Br $M_{4,5}$ $3d_{3/2,5/2}$ Core Level Absorptions during $Br_2$ C $^1\Pi_u$ $1_u$ State Dissociation via Extreme Ultraviolet Transient Absorption Spectroscopy

*John E. Beetar,*[1,†,‡] *Jen-Hao Ou,*[1,‡] *Yuki Kobayashi,*[2] *and Stephen R. Leone*[1,3,4,*]

[1]Department of Chemistry, University of California, Berkeley, California 94720, USA

[2]Department of Chemistry, University of Michigan Ann Arbor, Michigan 48109, USA

[3]Chemical Sciences Division, Lawrence Berkeley National Laboratory, Berkeley, California 94720, USA

[4]Department of Physics, University of California, Berkeley, California 94720, USA

[‡]These two authors contributed equally.

E-mail: srl@berkeley.edu

## ABSTRACT

The reported "dissociation times" for the $Br_2$ C ($^1\Pi_u$ $1_u$) state by various measurement methods differ widely across the literature (30 to 340 fs). We consider this issue by investigating attosecond extreme ultraviolet (XUV) transient absorption spectroscopy at the Br $M_{4,5}$ $3d_{3/2,5/2}$ edges (66 to 80 eV), tracking core-to-valence (3d → 4p) and core-to-Rydberg (3d → *n*s, *n*p, $n \geq 5$) transitions from the molecular to atomic limit. The progress of dissociation can be ascertained by the buildup of the atomic absorption in time. Notably, the measured rise times of the $3d_{5/2,\,3/2}$ → 4p transitions depend on the probed core level final state, 38 ± 1 and 20 ± 5 fs for $^2D_{5/2}$ and $^2D_{3/2}$ at 64.31 and 65.34 eV, respectively. Simulations by the nuclear time-dependent Schrödinger equation reproduce the rise-time difference of the 3d → 4p transitions, and the theory suggests several important factors. One is the transition dipole moments of each probe transition have different molecular and atomic values for $^2D_{5/2}$ versus $^2D_{3/2}$ that depend on the bond length. The other is the merger of multiple molecular absorptions into the same atomic absorption, creating multiple timescales even for a single probe transition. Unfortunately, the core-to-Rydberg absorptions did not allow accurate atomic Br buildup times to be extracted due to spectral overlaps with ground state bleaching, otherwise an even more comprehensive picture of the role of the probe state transition would be possible. This work shows that the measured probe signals accurately contain the dissociative wavepacket dynamics but also reveal how the specific probe transition affects the apparent progress toward dissociation with bond length. Such potential probe-transition-dependent effects need to be considered when interpreting measured signals and their timescales.

## 1. INTRODUCTION

Dissociation is a fundamental chemical reaction. While dissociation is simple to perceive as a phenomenon, there is not a once-and-for-all definition for dissociation time. In fact, Zewail and colleagues[1] pointed out that "dissociation time" could have more than one definition, when they studied the dissociation of ICN.

A good example of the subtlety of "dissociation times" is the dissociation of the $Br_2$ C $^1\Pi_u$ $1_u$ state into two ground state Br atoms (Nomenclature of $Br_2$ electronic states follows the literature). It is triggered by absorption of 400 nm light from the electronic ground state (X $^1\Sigma_g^+$) of $Br_2$ as

$$Br_2\ (X\ ^1\Sigma_g^+) + 400\text{ nm light} \rightarrow Br_2\ (C\ ^1\Pi_u\ 1_u) \rightarrow Br\ (^2P_{3/2}) + Br\ (^2P_{3/2}).$$

Over the past twenty years, this dissociation has been extensively studied by numerous time-resolved measurements.[2–9] Surprisingly, these measurements[2–9] report widely different dissociation times from 30 to 340 fs dependent on the various detection methods (Table 1).

In general, those detection methods measure two categories of detection, particles (photoelectrons[2–5,10] or photoions[7–9]) or light (emission[6,7]). With the detection of particles, time-resolved photoelectron spectroscopy (TRPES)[2–5,10] showed that the atomic signals appear around 40 fs[2–4] to 85 fs[5,10] after excitation. On the other hand, cold target recoil ion momentum spectroscopy (COLTRIMS)[8] found that the molecular signals persist up to 140 ± 15 fs. In addition, Coulomb explosion imaging (CEI)[9] recorded the rise time as 200 fs for $Br^+$ ions (probed by a 800 nm pulse) and 340 fs for $Br^{2+}$ ions (probed by a 90.6 eV pulse). With the detection of light emission, high harmonic spectroscopy (interferometry)[6,7] showed that the amplitude of emitted high harmonics converges to an asymptotic value around 300 fs, when the atomic character is considered to be fully established. Such differences in the dissociation time are intriguing, especially when the reactant $Br_2$ is a simple diatomic molecule.

In this work, we use attosecond XUV transient absorption spectroscopy[11–14] to track the dissociation of $Br_2$ in the C $^1\Pi_u$ $1_u$ state by core level absorptions. Specifically, the XUV absorption spectrum covers both the Br 3d core-to-valence and core-to-Rydberg absorptions at the same time. During the dissociation, the XUV absorption energy shifts because the energy gap between the C $^1\Pi_u$ $1_u$ and Br 3d core-excited states changes. By following the XUV absorption spectrum as it evolves from the molecular to the atomic limit, we map out the underlying change in the electronic structure from $Br_2$ molecules to Br atoms. We consider the dissociation to be completed based on spectroscopic identification[1] when the XUV absorption spectrum converges to that of the Br atoms, within the temporal and spectral resolution of the experiment.

Previously, the static Br 3d core-to-valence and core-to-Rydberg spectra were measured for either $Br_2$ molecules or Br atoms separately[15–21] (Table 2). The broadband time-resolved spectra connect these studies scattered throughout the literature with a full evolution from molecules to atoms. This unification demonstrates the unique strength of time-resolved absorption spectroscopy with broadband XUV light.

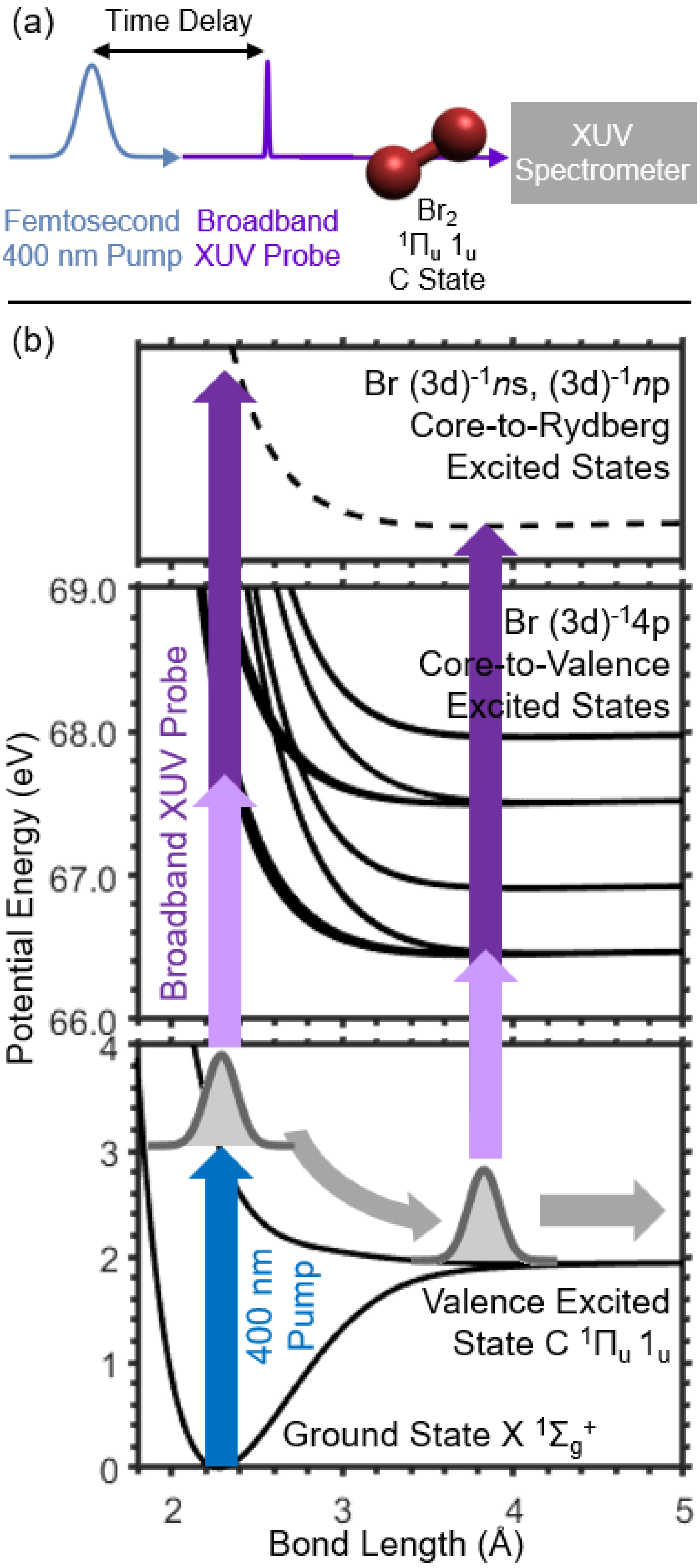

Figure 1. Experimental scheme. (a) Attosecond extreme ultraviolet (XUV) transient absorption spectroscopy on the dissociation of $Br_2$ molecules in the C $^1\Pi_u$ $1_u$ state. The dissociation is launched by femtosecond 400 nm pump pulses and monitored by the time-resolved absorption of broadband XUV probe pulses at the Br $M_{4,5}$ $3d_{3/2,5/2}$ edges (66 to 80 eV). (b) Femtosecond 400 nm pump pulses bring the $Br_2$ molecule from the electronic ground X $^1\Sigma_g^+$ state to the excited C $^1\Pi_u$ $1_u$ state, to initiate the dissociation. As the nuclear wavepacket travels on the C state, it is probed by the absorptions of XUV pulses to the Br core-to-valence $(3d)^{-1}4p$ and core-to-Rydberg $(3d)^{-1}ns$ and $(3d)^{-1}np$ ($n \geq 5$) excited states. Solid lines: Calculated potential energy curves (PEC) of the ground X, C and core-to-valence $(3d)^{-1}4p$ excited states. Dashed lines: Hypothetical PEC of core-to-Rydberg excited state for illustration purposes.

Table 1. Experimental time-resolved studies on the dissociation of $Br_2$ C $^1\Pi_u$ $1_u$ state

| Object of Detection: Experimental Technique | | | | |
|---|---|---|---|---|
| Timescale Reported (fs) | Temporal Resolution (fs) | Pump Pulse | Probe Pulse | Reference |
| **Particle (photoelectron): Time-resolved photoelectron spectroscopy (TRPES)** | | | | |
| 40 ± 14 ($Br^+$ $^3P_{2,1,0}$) | 300±14 | 400 nm, 80 fs | H17, 47 nm (26.4 eV), 290 fs | Nugent-Glandorf 2001 ([2]) |
| 45 ± 8 ($Br^+$ $^3P_{2,1,0}$) | 340±16 | 402.5 nm | H17, 47 nm (26.4 eV), 250 fs | Nugent-Glandorf 2002 ([3]) |
| 30 ± 19 ($Br^+$ $^3P_{2,1,0}$) | 203±17 | 402.5 nm | H19, 42 nm (29.45 eV), 190 fs | Nugent-Glandorf 2002 ([3]) |
| 40 ($Br^+$ $^3P_{2,1,0}$) | 230 | 402.5 nm, < $10^{12}$ W/cm$^2$ | H15, 53.7 nm, (23.1 eV) | Strasser 2007 ([4]) |
| 85 ± 15 ($Br^+$ $^3P_{2,1,0}$) | 135±5 | 395 nm, 60 fs, < $10^{12}$ W/cm$^2$ | H15, 23.5 eV, 120 fs | Wernet 2009 ([5]) |
| **Particle (photoion): Cold target recoil ion momentum spectroscopy (COLTRIMS)** | | | | |
| 140 ± 15 ($Br^+$) | | 400 nm, 40 fs, 2×$10^{11}$ W/cm$^2$ | 800 nm, 30 fs, 4×$10^{13}$ W/cm$^2$ | Li 2010 ([8]) |
| **Particle (photoion): Coulomb explosion imaging (CEI)** | | | | |
| 200 ($Br^+$) | | 400 nm, 100 fs, 1×$10^{12}$ W/cm$^2$ | 800 nm, 100 fs, 5×$10^{13}$ W/cm$^2$ | Rouzée 2013 ([9]) |
| 340 ($Br^{2+}$) | 300 | | 15.7 nm (90.6 eV), 50 fs, $10^{13}$ W/cm$^2$ | Rouzée 2013 ([9]) |
| Particle (photoion) and light (emission): High harmonic spectroscopy | | | | |
| 300 | 50 to 60 (H18) | 400 nm, $5\times10^{11}$ W/cm$^2$ | 800 nm, $1.5\times10^{14}$ W/cm$^2$ | Wörner 2010 ([7]) |
| Light (emission): High harmonic transient grating spectroscopy | | | | |
| 300 | 50 (H20) | 400 nm, $10^{12}$ W/cm$^2$ | 800 nm, $10^{14}$ W/cm$^2$ | Wörner 2010 ([6]) |
| Light (absorption): XUV transient absorption spectroscopy | | | | |
| 38 ± 1 Br 3d ($^2P_{3/2}$) → 4p ($^2D_{5/2}$) | 29±2 | 400 nm, 26 ± 1 fs, $5\times10^{12}$ W/cm$^2$ | 50-80 eV broadband | Present work |
| 20 ± 5 Br 3d ($^2P_{3/2}$) → 4p ($^2D_{3/2}$) | | | | |

Table 2. Measurements of Br core-to-valence (3d → 4p) and core-to-Rydberg (3d → *n*s, *n*p, $n \geq 5$) spectra of ground state $Br_2$ molecules and Br atoms in the literature.[15–21]

| System | Transitions | |
|---|---|---|
| | Core-to-valence<br>3d → 4p<br>(67-70 eV) | Core-to-Rydberg<br>3d → *n*s, *n*p, $n \geq 5$<br>(73-80 eV) |
| $Br_2$ molecule | EELS[15]<br>XUV-TAS[16] | XAS[21] |
| Br atom | EELS[15]<br>PES[18]<br>XAS[17,20]<br>XUV-TAS[16] | PES[19]<br>XAS[20] |

EELS: Electron impact Energy Loss Spectroscopy

PES: Photoelectron Spectroscopy

XAS: XUV/X-ray Absorption Spectroscopy (static)

XUV-TAS: XUV Transient Absorption Spectroscopy (time-resolved)

## 2. METHODS

The experimental and computational methods are described briefly here with details in the Supporting Information.

### 2.1 Experiments

The laser source is a Coherent Astrella USP (800 nm, 35 fs FWHM, 7 mJ, 1 kHz). The pump pulses (400 nm, 26 ± 1 fs FWHM) come from second harmonic generation (SHG) of 70% of the Astrella output (4.9 mJ) in a β-barium borate (BBO) crystal, the 400 nm light is focused through a neon-filled (1.7 bar) tube (1.37 m long), and the output is then compressed by 8 chirped mirrors. The XUV probe pulses (50-85 eV photon energy) come from high harmonic generation (HHG) in a neon gas cell with 800 nm driving field pulses. The driving field pulses are produced by taking 30% of the Astrella output (2.1 mJ), focusing it into a neon-filled (20 psi) stretched hollow core fiber (HCF) (1.5 m long, 400 μm inner diameter), and then temporally compressing the output by 16 chirped mirrors and a piece of 2 mm thick ADP crystal. The driving field is then removed by a 0.1 μm thick Al filter. At the sample cell, pump and probe beams recombine non-collinearly and interact with $Br_2$ vapor. After a 0.1 μm thick Zr filter, the XUV beam is detected by a homebuilt spectrograph. The time-resolved (transient) differential absorption $\Delta A(\omega, t)$ is calculated from the XUV absorption spectrum of $Br_2$ with and without pump pulses.

### 2.2 Computations

Potential energy curves (PECs) and transition dipole moments (TDMs) for valence and 3d-to-4p core-to-valence excited states are calculated with a developer version of GAMESS.[22] The nuclear wavepacket trajectory on the PEC of the C state is numerically solved from the nuclear time-dependent Schrödinger equation (TDSE) where the transition due to the pump pulse is explicitly considered. The time-resolved (transient) XUV absorption cross section[23] from the C state to core-to-valence excited states are calculated from the TDMs and wavepacket trajectory.

## 3. RESULTS AND DISCUSSION: Overview of both the Core-to-Valence and Core-to-Rydberg Absorption Spectrum

In this study, a 400 nm pump pulse excites $Br_2$ from the electronic ground state (X $^1\Sigma_g^+$) to the excited C $^1\Pi_u$ $1_u$ state, which causes the dissociation into two Br atoms in the electronic ground state ($^2P_{3/2}$).

$$Br_2\ (^1\Sigma_g^+\ X) + 400\ \text{nm light} \rightarrow Br_2\ (C\ ^1\Pi_u\ 1_u) \rightarrow Br\ (^2P_{3/2}) + Br\ (^2P_{3/2})$$

The above processes are monitored with XUV probe pulses via changes in the absorption spectrum from the molecular to atomic limit over time. Furthermore, due to the broad bandwidth of the XUV probe pulse, not only core-to-valence but also core-to-Rydberg absorptions are measured. Both absorptions will be overviewed in this section, then elaborated for core-to-valence absorptions in Section 4, then core-to-Rydberg absorptions in Section 5.

### 3.1 BEFORE Dissociation: Identification of $Br_2$ Molecules with Static Core-to-Valence and Core-to-Rydberg Absorption Spectrum

Before the pump pulse, the $Br_2$ molecules in the electronic ground state are identified with its XUV absorption spectrum (Figure 2 and Table 3). Both core-to-valence (3d → 4p, 67 to 70 eV) and core-to-Rydberg (3d → *n*s, *n*p ($n \geq 5$), 73 to 79 eV) transitions are observed together. For core-to-valence absorptions (Figure 2(a)), the $3d_{5/2,3/2} \rightarrow 4p\sigma^*$ spin-orbit doublet is at 68.10 and 69.13 eV. The spin-orbit splitting is 1.03 eV. For core-to-Rydberg absorptions (Figure 2(b)), a series of 3d → *n*s, *n*p ($n \geq 5$) absorption peaks are observed. Our results agree with the literature using electron impact energy loss spectroscopy[15] and photoelectron spectroscopy[21].

A clear XUV absorption spectrum of $Br_2$ molecules marks the starting point for the study of dissociation. As the dissociation is launched by the pump pulse, the absorption from $Br_2$ molecules decreases and is observed as ground state bleaching signals in the time-resolved XUV absorption spectrum in the next section.

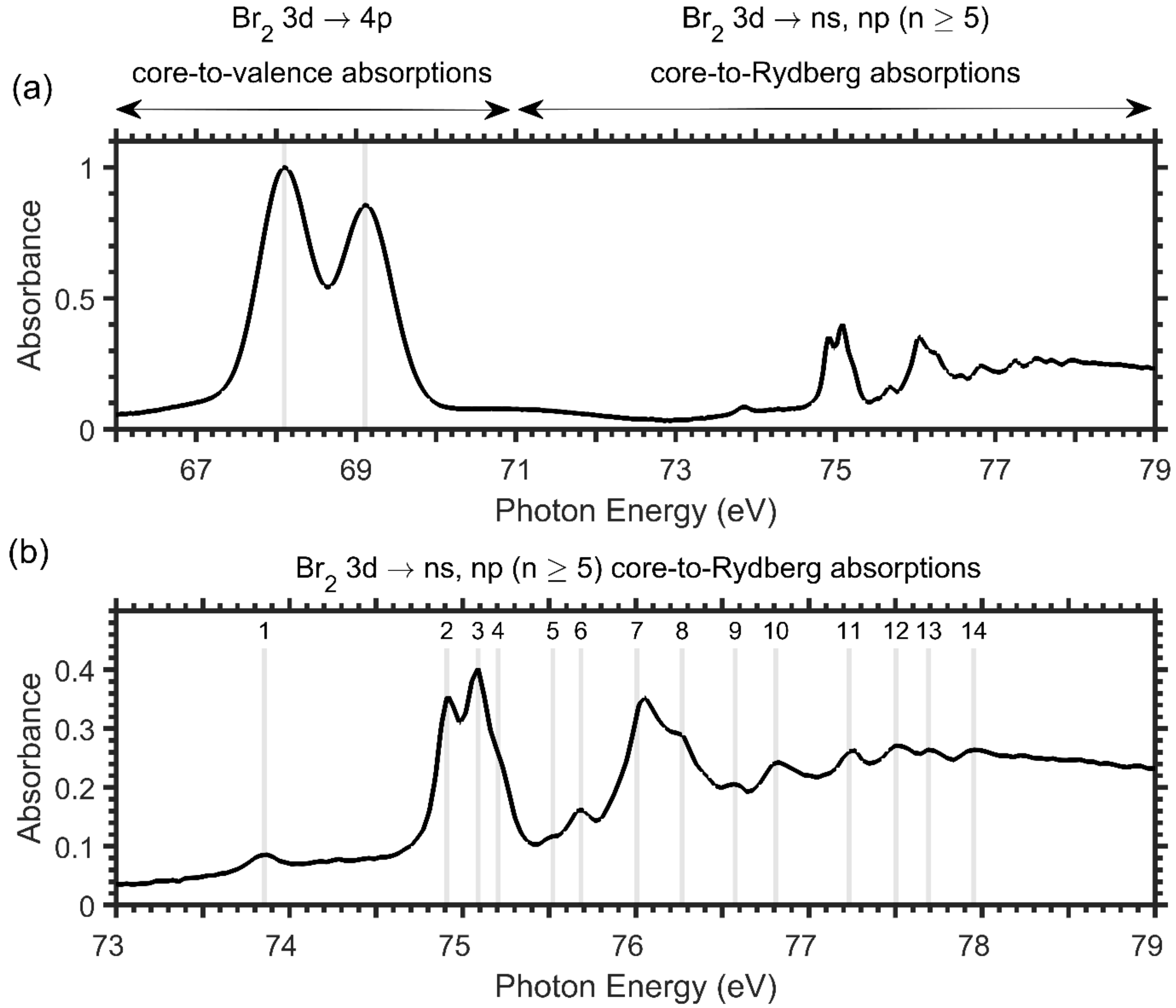

Figure 2. Static XUV absorption spectrum of $Br_2$ molecules at Br 3d M edge (a) 3d → 4p core-to-valence and 3d → *n*s, *n*p ($n \geq 5$) core-to-Rydberg absorptions. (b) 3d → *n*s, *n*p ($n \geq 5$) core-to-Rydberg absorptions. Assignments in Table 3.

Table 3. Br core-to-valence (3d → 4p) and core-to-Rydberg (3d → *n*s, *n*p ($n \geq 5$)) absorptions from $Br_2$ molecules in the electronic ground state. The corresponding spectrum is in Figure 2. Nomenclature of the core-to-Rydberg absorptions follows Ref. [21].

| Peak | Electronic Transitions | E (eV) This work[a] | E (eV) Ref. [15,b] | E (eV) Ref. [16,a] | E (eV) Ref. [21,c] |
|---|---|---|---|---|---|
| | core-to-valence | | | | |
| | $3d_{5/2} \rightarrow 4p\sigma^*$ | 68.10 | 68.189 | 68.3 | |
| | $3d_{3/2} \rightarrow 4p\sigma^*$ | 69.11 | 69.209 | 69.4 | |
| | core-to-Rydberg | | | | |
| 1 | 3d → 5s | 73.85 | | | |
| 2 | $3d_{\delta 5/2} \rightarrow 5p$ | 74.91 | 74.890 | | 74.946 |
| 3 | $3d_{\pi 3/2} \rightarrow 5p$ | 75.09 | 75.047 | | 75.112 |
| 4 | $3d_{\sigma 1/2} \rightarrow 5p$ | 75.21 | | | 75.238 |
| 5 | $3d_{\delta 5/2} \rightarrow 6s$ | 75.52 | | | 75.518 |
| 6 | $3d_{\pi 3/2} \rightarrow 6s$ | 75.68 | | | 75.672 |
| 7 | $3d_{\delta 3/2} \rightarrow 5p$ | 76.01 | 75.962 | | 76.007 |
| 8 | $3d_{\pi 3/2} \rightarrow 6p$ | 76.27 | | | 76.272 |
| 9 | $3d_{\delta 5/2} \rightarrow 7p$ | 76.57 | | | 76.555 |
| 10 | $3d_{\pi 1/2} \rightarrow 6s$ | 76.81 | | | 76.771 |
| 11 | $3d_{\delta 3/2} \rightarrow 6p$ | 77.23 | | | 77.178 |
| 12 | $3d_{\pi 1/2} \rightarrow 6p$ | 77.50 | | | 77.419 |
| 13 | $3d_{\delta 3/2} \rightarrow 7p$ | 77.69 | | | 77.613 |
| 14 | $3d_{\pi 1/2} \rightarrow 7p$ | 77.95 | | | 77.809 |

[a] XUV absorption spectroscopy

[b] Electron impact energy loss spectroscopy.[15] Energy resolution (FWHM) is 75 meV.

[c] Photoelectron spectroscopy.[21] Photon resolution is 50 meV at 90 eV photon energy.

### 3.2 DURING Dissociation: Overview of the Changes in Time-Resolved Core-to-Valence and Core-to-Rydberg Absorption Spectrum

The dissociation is monitored by the changes in the time-resolved XUV absorption spectrum. Figure 3(b) shows the time-resolved XUV differential absorption $\Delta A(t)$ spectrum as a function of the time delay between the 400 nm pump and XUV probe pulses. Positive time delays mean the 400 nm pump arrives before the XUV probe pulse. The differential absorption $\Delta A(t)$ signals represent the change in XUV absorption after the pump pulse. Positive and negative $\Delta A(t)$ signals correspond to new or increased absorptions and decreased absorptions, respectively. We will analyze the core-to-valence absorptions first in Section 4, then core-to-Rydberg absorptions in Section 5.

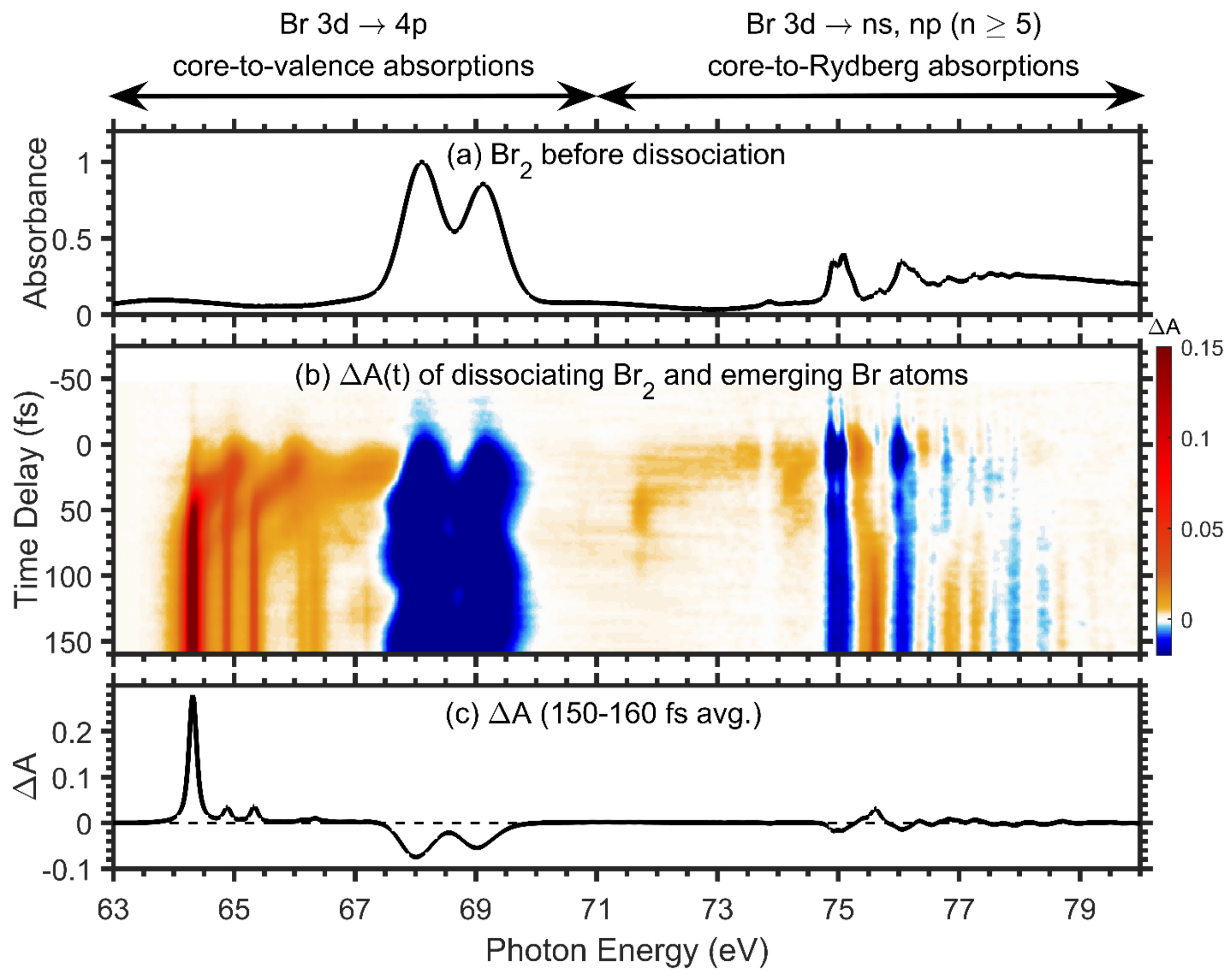

Figure 3. Measured time-resolved XUV core-to-valence (67-70 eV) and core-to-Rydberg (73-80 eV) absorption spectrum during the dissociation of the $Br_2$ C $^1\Pi_u$ $1_u$ state. (a) Static XUV absorption spectrum of the $Br_2$ molecules before dissociation. (b) Time-resolved XUV differential absorption spectrum ΔA(t) of the dissociating $Br_2$ molecule and resulting Br atoms, as a function of the time delay between the 400 nm pump and XUV probe pulses. Positive time delays mean the 400 nm pump pulse arrives before the XUV probe pulse. (c) Time average from 150 to 160 fs of the differential absorption signals ΔA(t) in (b). New absorptions from dissociated Br atoms lead to positive ΔA, and decreased absorptions from $Br_2$ molecules cause negative ΔA. Specifically, the $Br_2$ C state dissociates into two ground state Br atoms, which are probed by the atomic 3d $^2P_{3/2}$

$\rightarrow$ 4p $^2D_{5/2,\,3/2}$ absorptions at 64.31 and 65.34 eV, respectively. The other atomic 3d $^2P_{1/2} \rightarrow$ 4p $^2D_{3/2}$ absorption is also observed at 64.89 eV, but this signal originates from another excited state of $Br_2$ molecules (Supporting Information). Above 73 eV, ground state bleaching of $Br_2$ core-to-Rydberg absorptions overlap with the appearance of atomic Br core-to-Rydberg absorptions.

## 4. DURING Dissociation: Changes in Time-Resolved Br Core-to-Valence Absorption Spectrum

Our terminology on differential absorption signals follows the literature.[24] During the dissociation, several phenomena occur such as *ground state bleaching* of absorptions from $Br_2$ molecule, *excited state absorptions* from the $Br_2$ molecule pumped to the excited C $^1\Pi_u$ $1_u$ state, and *product absorptions* from the dissociating Br atoms, which are explained in the following subsections.

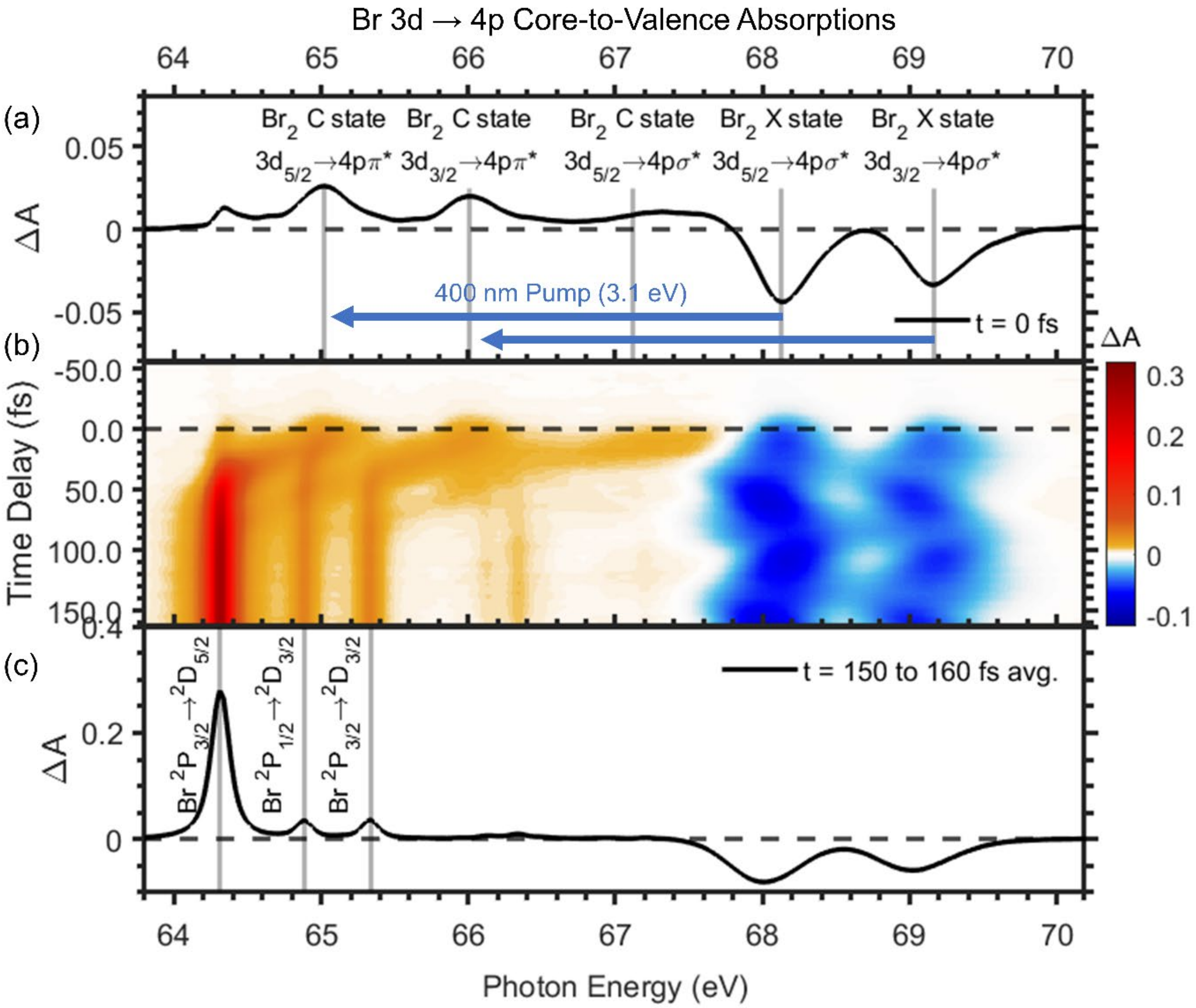

Figure 4. Measured time-resolved XUV core-to-valence absorption spectrum during the dissociation of the $Br_2$ C $^1\Pi_u$ $1_u$ state. (a) The differential absorption spectrum at time zero ΔA(t = 0). The negative signals around 67.5-69.5 eV are the ground state bleaching of the $3d_{5/2,3/2} \rightarrow 4p\sigma^*$ doublet of $Br_2$ molecules. (b) Time-resolved differential absorption spectrum ΔA(t) of the dissociating $Br_2$ molecule and resulting Br atoms, as a function of the time delay between the 400 nm pump and XUV probe pulses. (c) Time average from 150 to 160 fs of the differential absorption signals ΔA(t) in (b).

## 4.1 Ground State Bleaching: Decrease of $Br_2$ Molecules in The Electronic Ground State (X)

The spectral region for core-to-valence absorptions is expanded in Figure 4. Once a fraction of the $Br_2$ molecules are excited by the pump pulse, the XUV absorptions of $Br_2$ molecules in the electronic ground state decreases, shown as the negative ΔA(t) signals (so-called ground state bleaching[24]) in Figure 4(b). The decrease of the $3d_{5/2,3/2} \rightarrow 4p\sigma^*$ absorption doublet leads to the broad negative signals around 67.5-69.5 eV, which are also seen in the snapshot at time zero ΔA(t = 0) in Figure 4(a) and the time average of ΔA(t) from 150 to 160 fs in Figure 4c. Additionally, the oscillation of ground state bleaching signals encode the dynamics of the coherent vibrational wavepacket[16,25–28] in the electronic ground state of $Br_2$ molecules (Supporting Information).

## 4.2 Excited State Absorption: $Br_2$ Molecules Pumped to the Electronic Excited State (C $^1\Pi_u$ $1_u$)

As the $Br_2$ molecules are pumped to the excited state (C $^1\Pi_u$ $1_u$), new XUV absorptions from the C state appear as positive ΔA(t) signals in Figure 4. In particular, a new absorption doublet arises at $t = 0$ due to the $3d_{5/2,3/2} \rightarrow 4p\pi^*$ absorption at 65.02 and 66.01 eV (a). This new absorption doublet of the C state $Br_2$ molecule is about 3 eV below the original $3d_{5/2,3/2} \rightarrow 4p\sigma^*$ absorption doublet (68.12 and 69.16 eV) of ground state $Br_2$ molecules. Physically, this shift corresponds to the energy gap between the excited C and ground X states, equivalent to the 3.1 eV photon energy of the 400 nm pump pulse. In addition, the $3d_{5/2} \rightarrow 4p\sigma^*$ absorption from the C state occurs around 67 eV, about 2 eV higher than the $3d_{5/2} \rightarrow 4p\pi^*$ absorption at 65.02 eV. This difference agrees with the 2.1 eV gap between orbital energy of $Br_2$ $4p\pi^*$ and $4p\sigma^*$ orbitals.

**4.3 From Molecules to Atoms: Mapping the Motion of the Nuclear Wavepacket on the C State Potential Energy Curve (PEC) By Core-to-Valence Absorptions**

After time zero, the C state $Br_2$ molecule dissociates into two ground state Br atoms

$$Br_2\ (C\ ^1\Pi_u\ 1_u) \rightarrow Br\ (^2P_{3/2}) + Br\ (^2P_{3/2})$$

As a result, the molecular $3d_{3/2,5/2} \rightarrow 4p\pi^*$ absorptions shift toward lower photon energy in Figure 4(b), and the absorptions gradually converge to the atomic 3d $^2P_{3/2} \rightarrow$ 4p $^2D_{5/2,\ 3/2}$ transitions at 64.31 and 65.34 eV, respectively. The atomic absorption peaks are clearly identified with the time average of $\Delta A(t)$ from 150 to 160 fs in Figure 4(c) and Table 4.

Such a down shift of the XUV absorption energy equals the decreasing energy gap between the C state and the $(3d)^{-1}4p$ core-to-valence excited state (white dash-dot line, Figure 5(b)), as the nuclear wavepacket descends along the C state PEC (Figure 5(a)). Hence, the motion of the nuclear wavepacket on the PEC is intuitively visualized in the time-resolved XUV absorption spectrum (Figure 5(b)).[29]

In addition to the atomic 3d $^2P_{3/2} \rightarrow$ 4p $^2D_{5/2,\ 3/2}$ absorptions, a weaker atomic 3d $^2P_{1/2} \rightarrow$ 4p $^2D_{3/2}$ absorption is also observed at 64.89 eV in Figure 4(c). This absorption from the atomic $^2P_{1/2}$ excited state is not a dissociation product of the C state but other valence excited states. Our simulations suggest that a possible state is the Z ($J = 2$) state (Supporting Information).

Table 4. Br core-to-valence 3d → 4p absorptions of dissociated Br atoms.

| Atomic Transition | E (eV) This work | E (eV) Ref. [20] | E (eV) Ref. [17] | E (eV) Ref. [16] | E (eV) Ref. [18] |
|---|---|---|---|---|---|
| $^2P_{3/2} \rightarrow 4p\ ^2D_{5/2}$ | 64.31 | 64.38 | 64.38 | 64.4 | 64.54 |
| $^2P_{1/2} \rightarrow 4p\ ^2D_{3/2}$ | 64.89 | 64.97 | 64.97 | 65.1 | 65.13 |
| $^2P_{3/2} \rightarrow 4p\ ^2D_{3/2}$ | 65.34 | 65.43 | 65.43 | 65.4 | 65.58 |

### 4.4 Product Absorption: Buildup Signal of Dissociated Br Atoms and its Fitted Rise Time

The lineout in time for the atomic 3d $^2P_{3/2} \rightarrow$ 4p $^2D_{5/2,\ 3/2}$ absorptions at 64.31 and 65.34 eV is plotted in Figure 5(d, c). Qualitatively, as the $Br_2$ molecule dissociates, the atomic absorption will increase and converge to a plateau. Quantitatively, to determine the rise time, the atomic differential absorption signal $\Delta A(\mathrm{Br}; t)$ is fitted to a sigmoid function[30,2,3,5,9,10] as

$$\Delta A(\mathrm{Br}; t) = a\, \Phi(t - t_{\mathrm{rise}}) \tag{1}$$

where $a$ is the amplitude, $\Phi(t)$ is the cumulative distribution function (CDF) of the pump pulse shape $f(t)$

$$\Phi(t) = \int_{-\infty}^{t} f(t')dt' \tag{2}$$

and $t_{\mathrm{rise}}$ is the time constant for a shift of the CDF function in time, called the apparent rise time. For the atomic 3d $^2P_{3/2} \rightarrow$ 4p $^2D_{5/2,\ 3/2}$ transitions (64.31 and 65.34 eV, Figure 5(d, c)), the fitted values of $t_{\mathrm{rise}}$ are 38 ± 1 and 20 ± 5 fs (Here the best estimate ± uncertainty is reported as the mean ± one standard deviation[31] of fitted values for each of 93 measurements.).

The physical meaning of this fit function has been elaborated in Zewail's landmark work (Section II.B and Appendix therein)[30] and is briefly explained here. This fit function implies that

(1) The absorption of pump pulse from the ground to the C state is assumed instantaneous.[30,32,33] That is, the response function of the molecules or atoms is assumed as a step function (Equation A1 in Ref. [30] and Figure 6(a) in Ref. [1]).

(2) The atomic XUV absorption signals are proportional to the populations pumped to the C state and emerge later with a shift in time by $t_{\mathrm{rise}}$ as the molecules in the C state gradually dissociate.[2,3,5,9,10] The shift in time by $t_{\mathrm{rise}}$ empirically describes the time scale of the buildup of dissociated atomic signals. It is an apparent rise time because it corresponds to the total measured signal at that photon energy, but not necessarily a single transition if there are degeneracies.

(3) In this work, the pump pulse shape $f(t)$ is the temporal intensity envelope measured from SD-FROG (Figure S3(d), Supporting Information), so the realistic effect of finite pump pulse duration and satellite pulses are explicitly considered. In the literature, an ideal Gaussian pulse shape is often used, where the Gaussian CDF equals an error function.[2,3,5,9,10]

(4) If a more sophisticated kinetics modeling is needed, a system of rate equations can be used.[30]

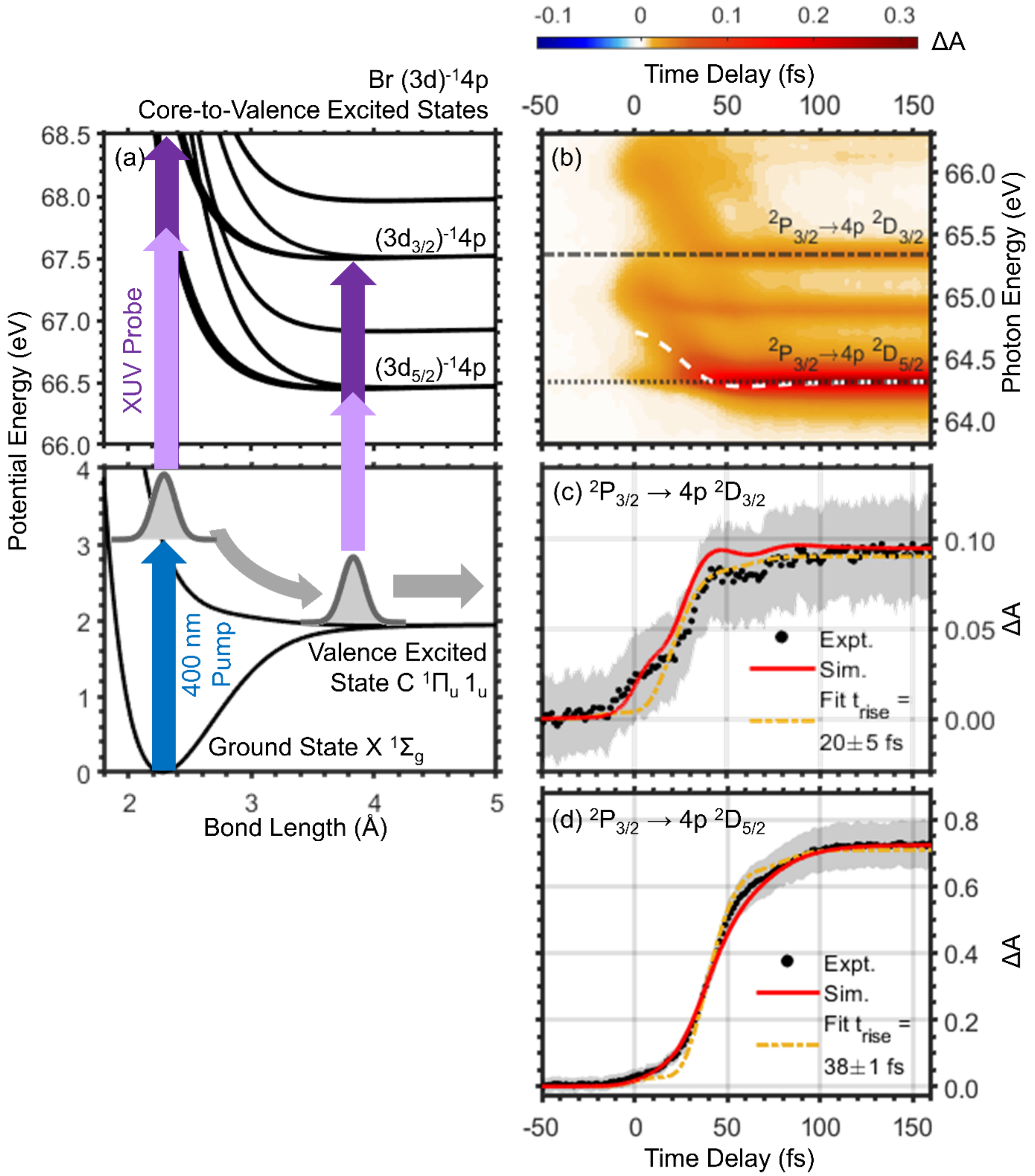

Figure 5. Launch and detection of the nuclear wavepacket and dissociation in the valence excited state C. (a) Potential energy curves. (b) Measured time-resolved XUV differential absorption spectrum records the motion of wavepacket on the PEC of C state and the rise of Br atomic absorptions. White dashed line: calculated energy gap between the C state and the Br $(3d_{5/2})^{-1}4p$

core-to-valence excited state. (d, c) Lineout in time of (b) at 64.31 and 65.34 eV for atomic Br 3d $^2P_{3/2} \rightarrow$ 4p $^2D_{5/2, 3/2}$ absorptions. Black dot and gray shade: the best estimate and the uncertainty of our measured ΔA(t) values (mean and one standard deviation[31] of 93 measurements of ΔA(t) signals). Red solid line: corresponding lineout of theoretically computed (GAMESS) and simulated time-resolved XUV absorption cross section from C state. Orange dash-dotted line: Fit by a Sigmoid function (Equation 1).

### 4.5 Potential Causes for the Different Rise Times: Comparison between Experiment, Simulation and Theory

#### 4.5.1 Comparison Between Experimental and Simulated Time-Resolved XUV Absorption Spectrum and the Underlying Theory

To understand the cause of the two different rise times, we compare the experimental time-resolved absorption spectrum to simulation and the underlying theory. The experimental time-resolved differential absorption spectrum (Figure 6(a)) is compared to the simulated time-resolved total absorption cross section of the C state to all core-to-valence excited states (Figure 6(b)). The photon energies of 64.31 and 65.34 eV are marked with the black dotted and dash-dotted lines for atomic $^2P_{3/2} \rightarrow$ 4p $^2D_{5/2}$ and $^2D_{3/2}$ absorptions, and lineouts in time are in Figure 5(d and c), respectively. For the lineout in time for each absorption in Figure 5(d or c), the measured (black dots) and simulated (red solid line) ΔA signals generally match. As a result, the simulations can capture the main features of the experiments. Next, we will analyze the components of the simulations and identify the potential cause of the different rise times.

Our simulations are based on a quasi-classical theory for transient absorption spectroscopy (Supporting Information).[23] According to the theory, the measured time-resolved *total* absorption cross section $\sigma_{\mathrm{C}}^{\mathrm{total}}(\omega, t)$ (Figure 6(b)) at photon energy *ħω* from the initial state (here the C state) to all final states ($3d^{-1}4p$ core-to-valence excited states) has two components: (1) the bond-length (*R*) dependent *total* absorption strength of the C state $\sigma_{\mathrm{C}}^{\mathrm{total}}(\omega, R)$ (Figure 6(c)) and (2) the wavepacket trajectory on C state PEC $|\Psi_{\mathrm{C}}(R,t)|^2$ (Figure 6(d)) as

$$\sigma_{\mathrm{C}}^{\mathrm{total}}(\omega, t) = \frac{4\pi\omega}{c} \int \sigma_{\mathrm{C}}^{\mathrm{total}}(\omega, R) |\Psi_{\mathrm{C}}(R,t)|^2 dR \qquad (3)$$

As a result, the measured time-resolved *total* absorption cross section $\sigma_{\mathrm{C}}^{\mathrm{total}}(\omega, t)$ shows not only how the wavepacket moves in time *t*, but also how the bond-length dependent *total* absorption strength $\sigma_{\mathrm{C}}^{\mathrm{total}}(\omega, R)$ changes with increasing bond length *R*. The wavepacket trajectory will be discussed later in Section 4.6.

### 4.5.2 The Difference between the Two Atomic Absorptions ($^2P_{3/2} \rightarrow 4p\ ^2D_{5/2}$ and $^2D_{3/2}$)

We first consider the simulated bond-length dependent *total* absorption strength of the C state $\sigma_{\mathrm{C}}^{\mathrm{total}}(\omega, R)$ in Figure 6(c). For the corresponding lineouts in Figure 6(e), the lineout of $\sigma_{\mathrm{C}}^{\mathrm{total}}(\omega, R)$ at 65.34 eV ($3d\ ^2P_{3/2} \rightarrow 4p\ ^2D_{3/2}$) (black dash-dotted line) converges to a plateau at a shorter bond length *R* than that at 64.31 eV ($3d\ ^2P_{3/2} \rightarrow 4p\ ^2D_{5/2}$) (black dotted line). This agrees with the above result that the rise time of $3d\ ^2P_{3/2} \rightarrow 4p\ ^2D_{3/2}$ absorption (20 ± 5 fs) is faster than that of $3d\ ^2P_{3/2} \rightarrow 4p\ ^2D_{5/2}$ absorption (38 ± 1 fs) (Figure 5).

### 4.5.3 Subtlety in the Atomic Absorption ($^2P_{3/2} \rightarrow 4p\ ^2D_{5/2}$): Merger of Two Branches of Molecular Absorptions (3d → 4pπ* and 4pσ*)

Furthermore, at 64.31 eV (atomic $3d\ ^2P_{3/2} \rightarrow 4p\ ^2D_{5/2}$ absorption), the lineout of $\sigma_{\mathrm{C}}^{\mathrm{total}}(\omega, R)$ shows a change in slope around 3 Å (Figure 6(e)). According to the full $\sigma_{\mathrm{C}}^{\mathrm{total}}(\omega, R)$ (Figure 6(c)), this atomic $3d\ ^2P_{3/2} \rightarrow 4p\ ^2D_{5/2}$ absorption at 64.31 eV originates from two branches of molecular absorptions from the C state (3d → 4pπ* and 4pσ*), which start separately around 64.9 and 67.1 eV at the equilibrium bond length $R_{eq}$ and merge together around 3 Å. (Their energy difference at $R_{eq}$ matches the 2.1 eV gap[34] between the orbital energy of 4pπ* and 4pσ* at $R_{eq}$.) As a result, for the lineout at 64.31 eV (Figure 6(e)), the molecular 3d → 4pπ* branch is the main contribution before 3 Å. After 3 Å, the other molecular 3d → 4pσ* branch blends in. Such a slope change is also observed in the lineout of the measured spectrum at 64.31 eV (Figure 5(d)), of which the slope changes around 50 fs. This phenomenon reveals the subtlety in interpreting the absorption spectrum due to spectral overlap. If multiple molecular absorptions converge to the same atomic absorptions, a single lineout at the photon energy for that atomic absorption actually contains the contributions from different molecular absorptions, complicating the observed timescale.

### 4.5.4 Summary: Two Factors Affecting the Lineout in Photon Energy of Measured Signals

In fact, the bond-length dependent *total* absorption strength of C state $\sigma_{\mathrm{C}}^{\mathrm{total}}(\omega, R)$ (Figure 6(c)) is the sum of all absorptions from the C state to each core-to-valence excited state (indexed by α) $\sigma_{\mathrm{C}\to\alpha}(\omega, R)$ as

$$\begin{aligned} \sigma_{\mathrm{C}}^{\mathrm{total}}(\omega, R) &= \sum_{\alpha} \sigma_{\mathrm{C}\to\alpha}(\omega, R) \\ \sigma_{\mathrm{C}\to\alpha}(\omega, R) &= |\mu_{\mathrm{C}\to\alpha}(R)|^2 g_{\mathrm{C}\to\alpha}(R, \omega). \end{aligned} \tag{4}$$

where $\mu_{\mathrm{C}\to\alpha}(\omega, R)$ is the transition dipole moment (TDM) from the C state to the *α*-th core-to-valence excited state and $g_{\mathrm{C}\to\alpha}(\omega, R)$ is a Lorentzian lineshape function. To summarize, two factors cause the rise times to differ for the lineouts of measured signals at two photon energies. One is how each absorption and its $\mu_{\mathrm{C}\to\alpha}(\omega, R)$ value change differently as the bond length increases. The other is the overlap of multiple degenerate absorptions at one XUV photon energy. For instance, when several molecular absorptions merge to the same atomic absorption.

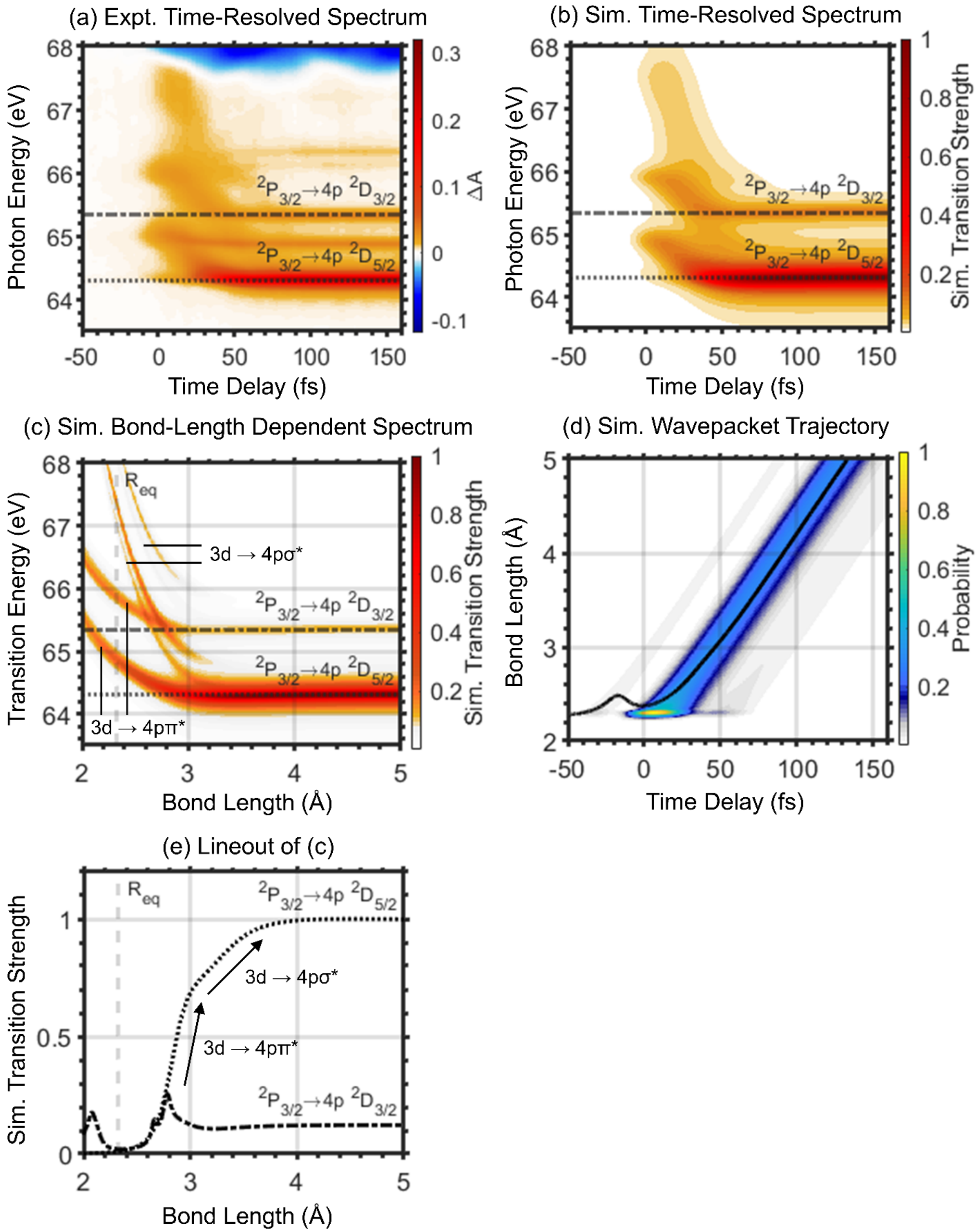

Figure 6. Comparison between experimental and simulated time-resolved XUV core-to-valence absorption spectrum and their components. The dissociated atoms are probed by the atomic 3d $^2P_{3/2} \rightarrow$ 4p $^2D_{5/2}$ and $^2D_{3/2}$ absorptions at 64.31 and 65.34 eV (black dotted line and black dash-dotted line). (a) Experimental time-resolved differential absorption spectrum. (b) Simulated time-resolved total absorption cross section $\sigma_{\mathrm{C}}^{\mathrm{total}}(\omega, t)$ of C state. It is the integral over bond length of (c) simulated bond-length dependent total XUV absorption cross section of C state $\sigma_{\mathrm{C}}^{\mathrm{total}}(\omega, R)$ and (d) simulated wavepacket trajectory on C state PEC $|\Psi_{\mathrm{C}}(R, t)|^2$. (e) Lineout of (c).

**4.6 Bond Length vs Time: General Agreement across the Literature**

The trajectory of the nuclear wavepacket on the PEC of the C state is calculated by numerically solving the nuclear TDSE (Figure 7(a)), where the temporal intensity envelope of the pump pulse measured by SD-FROG is used (Figure 7b and Supporting Information Figure S3). The position expectation value of the wavepacket trajectory links the bond length and the time during the dissociation. The agreement with the literature[2,3,5,6,8] (Figure 7 and Table 5) confirms the validity of the nuclear TDSE simulations. More importantly, even though the measured "dissociation times" in the literature could be different by an order of magnitude, how the bond length increases in time is overall consistent across different studies. Reported simulations of the nuclear motion (either classical or quantum wavepacket)[2,3,5,6,8] and this work generally agree that the bond length of $Br_2$ is double $R_{eq}$ (2.28105 Å)[35] or around 100 to 120 fs.

## 4.7 When does a chemical bond break?

According to the IUPAC Compendium of Chemical Terminology[36] (the Gold Book), dissociation is defined as "The separation of a molecular entity into two or more molecular entities." Here for $Br_2$, the separation for a diatomic molecule into two atoms can be quantified by the bond length. With the consensus on how bond length increases in time among the literature[2,3,5,6,8] as discussed above, the next question is: At which bond length should the bonding be considered vanished?

To answer this question, various detection methods and criteria have been used (Table 1), leading to different "dissociation times" for the $Br_2$ C $^1\Pi_u$ $1_u$ state. Even with the same experimental technique, the timescale may differ depending on the specific transitions or how the signal is selected and interpreted. There have been extensive discussions[4–6,8,10] on how the timescales from different experimental techniques may be reconciled or understood. It is important to recognize what and how a phenomenon is observed when one interprets its meaning and connects its corresponding timescale to dissociation.

For future studies, ultrafast electron diffraction (UED) may be used to measure how the bond length changes in time during the dissociation and compare that with existing simulations in the literature. As the temporal resolution of UED (about a hundred fs)[37,38] may be close (in order of magnitude) to the time scale of bond-length separation (double $R_{eq}$ around 100 to 120 fs), the temporal smearing of the observation will need to be tested and considered.

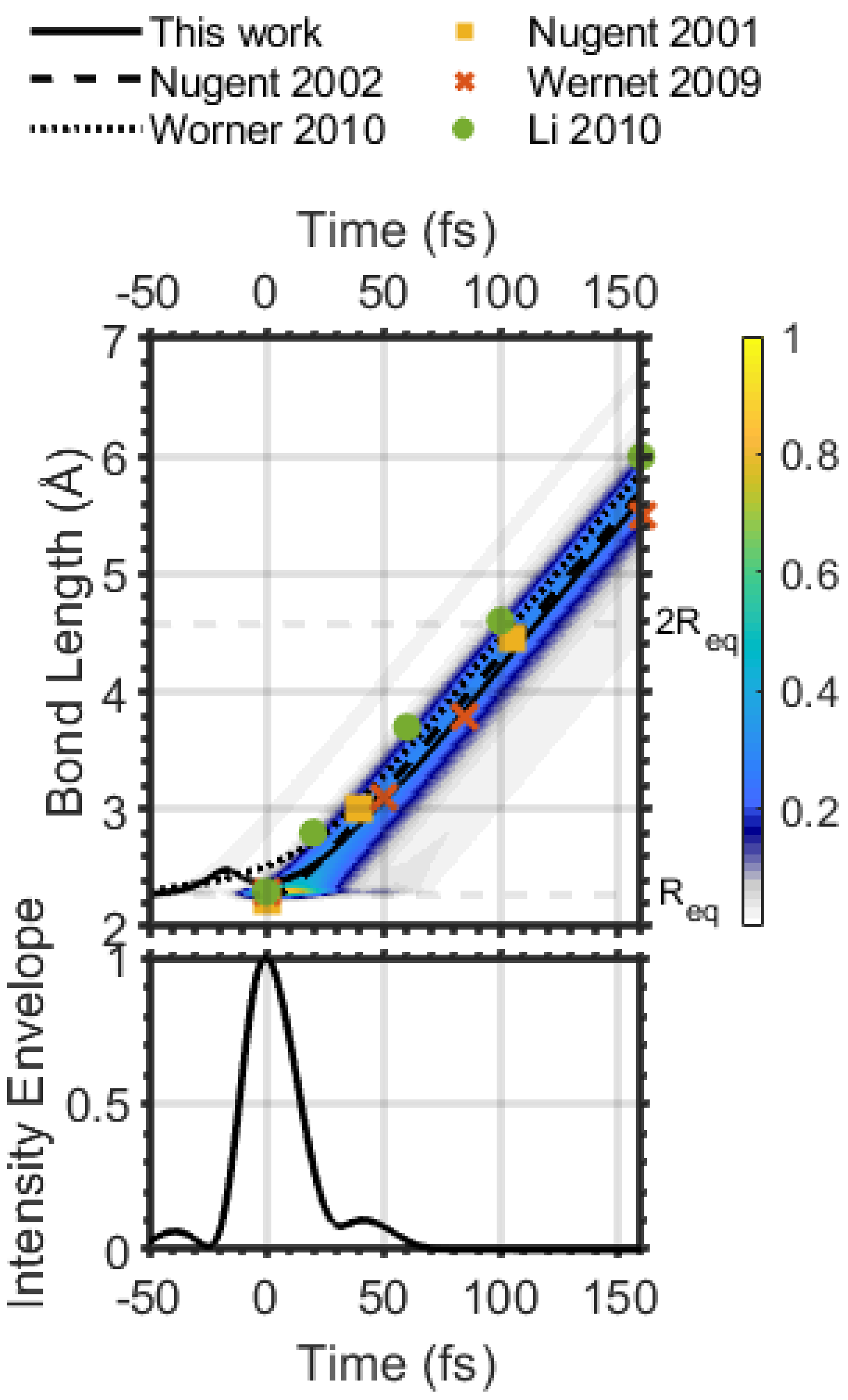

Figure 7. (a) Calculated nuclear wavepacket trajectory and its position expectation value (black solid line) on the PEC of excited state C from numerically solving the nuclear TDSE and comparison with the literature. Gray dashed lines: one and double equilibrium bond length $R_{eq}$ = 2.28105 Å.[35] Black dashed line: Adapted with permission from Fig. 5(b) in Ref. [3]. Copyright 2002 AIP Publishing. Black dotted line: Adapted with permission from Fig. S2 in Ref. [6]. Copyright 2010 Springer Nature. Yellow square: Ref. [2]. Red cross: Ref. [5]. Green circle: Ref. [8]. (b) Pump pulse shape (the temporal intensity envelope measured by SD-FROG) used in the TDSE. The effect of small pre- and post-pulses are seen as the additional weaker trajectories before and after time zero.

Table 5. Calculated bond length ($R$) as a function of time ($t$) for dissociation of $Br_2$ molecules on the C state PEC.

| Reference | $t$ (fs) | $R$ (Å) | $R/R_{eq}$ | $R$ (Å) This work |
|---|---|---|---|---|
| This work | 0 | 2.28 ($R_{eq}$) | 1.00 | |
| | 38 ± 1 | 2.91 ± 0.02 | 1.25 | |
| | 118.2 | 4.64 | 2.00 | |
| | 166.8 | 5.80 | 2.50 | |
| Nugent-Glandorf et al. 2001[2] | 0 | 2.23 ($R_{eq}$) | 1.00 | 2.32 |
| | 40 | 3.00 | 1.35 | 2.87 |
| | 105 | 4.46 | 2.00 | 4.32 |
| Wernet et al. 2009[5] | 0 | 2.3 ($R_{eq}$) | 1.0 | 2.32 |
| | 50 ± 15 | 3.1 ± 0.3 | 1.3 | 3.07 ± 0.20 |
| | 85 ± 15 | 3.8 ± 0.3 | 1.6 | 3.85 ± 0.24 |
| | 160 | 5.5 | 2.4 | 5.64 |
| Li et al. 2010[8] | 0 | 2.3 ($R_{eq}$) | 1.0 | 2.32 |
| | 20 | 2.8 | 1.2 | 2.37 |
| | 60 | 3.7 | 1.6 | 2.52 |
| | 100 | 4.6 | 2.0 | 4.21 |
| | 160 | 6.0 | 2.6 | 5.64 |

## 5. DURING Dissociation: Changes in Time-Resolved Br Core-to-Rydberg Absorption Spectrum

### 5.1 Challenges to Interpret the Core-to-Rydberg Absorption Spectrum

So far, we have analyzed the core-to-valence absorptions around 63-70 eV. The core-to-Rydberg absorptions (73-80 eV) are more challenging to interpret, because several different features overlap in the near spectral region. The most prominent one is the overlap of the ground state bleaching of $Br_2$ molecular absorptions and the emerging Br atomic absorptions as explained below.

### 5.2 Ground State Bleaching: Decrease of $Br_2$ Molecules in the Electronic Ground State (X)

The region of core-to-Rydberg absorptions (73.5-80 eV) is expanded in Figure 8. The decreases in molecular $Br_2$ $3d_{\delta 5/2} \rightarrow 5p$, $3d_{\pi 3/2} \rightarrow 5p$, and $3d_{\delta 3/2} \rightarrow 5p$ absorptions result in three noticeable negative $\Delta A(t)$ signals around 74.91, 75.09, and 76.02 eV, respectively (Peaks 2, 3 and 7 in Figure 8(a) and Table 3).

The lineouts in time for these three transitions are plotted in Figure 9. Ground state bleaching in core-to-Rydberg absorptions here shows several differences to the core-to-valence ones. First is a strong negative suppression signal around time zero. This signal resembles the shape of the pump pulse and may be a pump-induced effect on the XUV absorption to the core-to-Rydberg excited states. A related phenomenon was reported for the iodine 4d → 6p core-to-Rydberg absorptions in ICl.[39] Second, no noticeable periodic temporal oscillations are observed in the core-to-Rydberg absorptions.

In addition to these three transitions, the bleaching in other core-to-Rydberg absorptions is also partially observed as the negative $\Delta A(t)$ signals in Figure 8(b). However, they are obscured by the overlap of atomic Br core-to-Rydberg absorptions in the same spectral region, as identified in the next section.

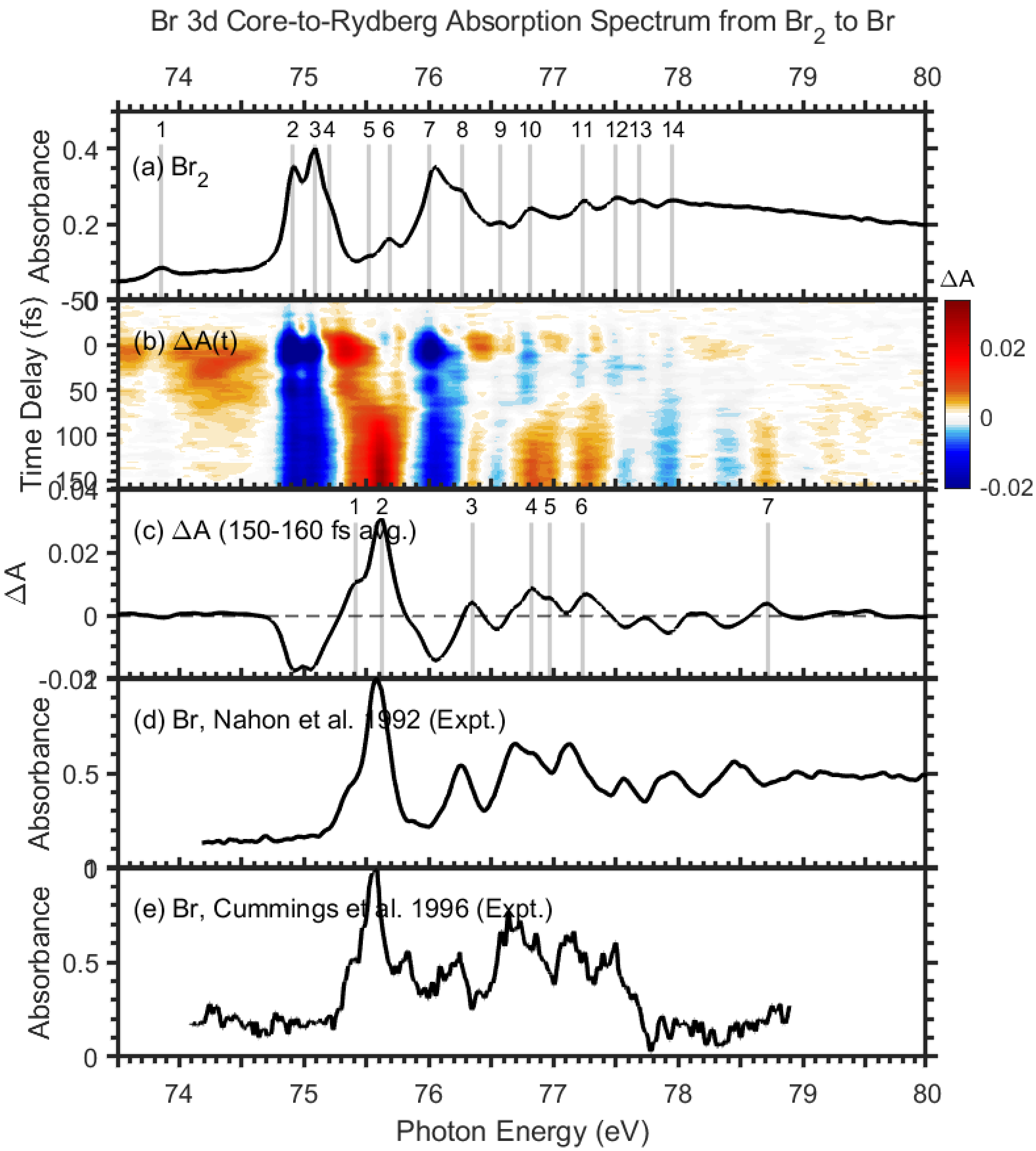

Figure 8. Static and time-resolved Br 3d Core-to-Rydberg XUV absorption spectrum of $Br_2$ molecules and Br atoms. (a) Static absorption spectrum of the $Br_2$ molecules before dissociation, as assigned earlier in Figure 2(b) and Table 3. (b) Time-resolved differential absorption spectrum $\Delta A(t)$ of the dissociating $Br_2$ molecule and resulting Br atoms, as a function of the time delay between the visible pump and XUV probe pulses. (c) Time average from 150 to 160 fs of the differential absorption signals ΔA(t) in Figure 8(b). New absorptions from dissociated Br atoms lead to positive $\Delta A$ (assigned in Table 6), while decreased absorptions from original $Br_2$ molecules lead to negative $\Delta A$. (d) Measured static absorption spectrum of Br atoms by photoelectron spectroscopy (average photon-energy resolution 0.18 eV). Adapted with permission from Ref. [19]. Copyright 1992 American Physical Society. (e) Measured static absorption spectrum of Br atoms by absorption spectroscopy (measurement accuracy ±0.03 eV). Adapted with permission from Ref.[20]. Copyright 1996 American Physical Society.

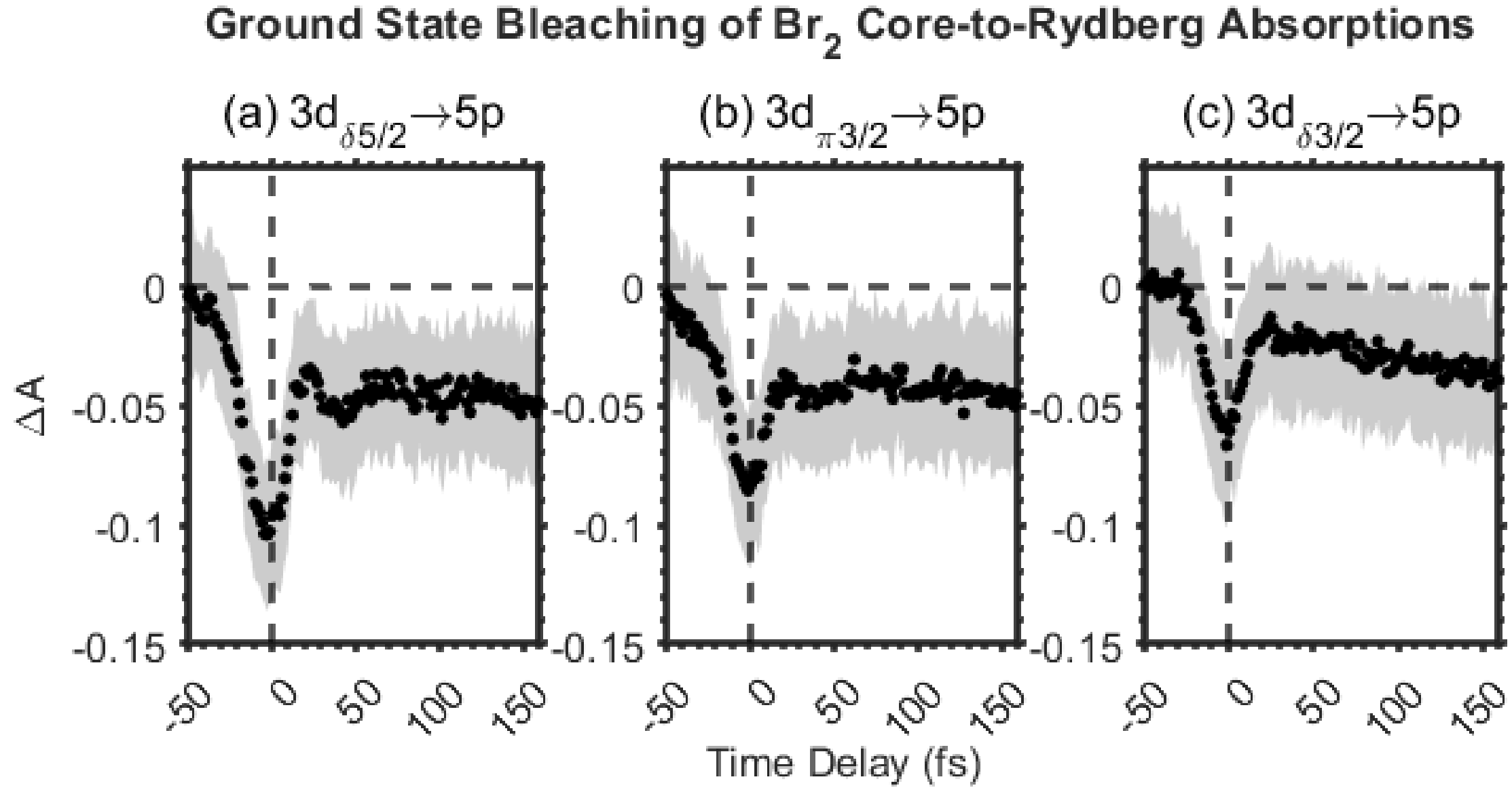

Figure 9. Differential absorption ΔA(t) signal as a function of time delay for selected $Br_2$ core-to-Rydberg absorptions. (a, b, c) Peaks 2, 3 and 7 in Figure 8(a) and Table 3. Black dot and gray

shade: best estimate (mean) and uncertainty (one standard deviation)[31] of 93 measurements. In addition to the ground state bleaching, a strong negative suppression signal is seen around time zero, potentially due to pump-induced effects.[39]

### 5.3 Product Absorption: Identification of Dissociated Br Atoms by Core-To-Rydberg Absorption Spectrum at Long Time Delay

New absorptions from dissociated Br atoms lead to positive ΔA(t) signals in Figure 8(b, c). The positive ΔA peaks in Figure 8(c) are compared to the static absorption spectra of Br atoms from the literature[19,20] in Figure 8(d, e). Generally, the overall shapes of the spectra agree with each other, and several noticeable peaks in Figure 8(c) are identified and assigned (Table 6). In addition, numerous other atomic Br core-to-Rydberg absorptions exist in this spectral region, and previous researchers pointed out the complexity of such atomic halogen core-to-Rydberg absorptions.[19] Readers are referred to the specific literature for detailed assignments.[19,20,40]

### 5.4 Product Absorption: Time Evolution of the Atomic Br Core-to-Rydberg Absorptions

The lineout in time for the identified positive ΔA peaks in Figure 8(c) are plotted in Figure 10. Ideally, the signal from dissociated Br atoms should rise around time zero and converge to a positive plateau at long time delay, similar to the atomic Br core-to-valence signals (Figure 5(c, d)). However, multiple additional factors complicate the analysis.

First is the overlap of various atomic Br and molecular $Br_2$ core-to-Rydberg absorptions (Figure 8). If the ground state bleaching of molecular absorptions is stronger than emerging atomic absorptions, the measured total ΔA signal at that photon energy may be dragged down and appear as a negative signal around time zero (Figure 10(d, e, f)). Such an overlap issue was not significant for core-to-valence absorptions, because major atomic (64.0-65.5 eV) and molecular (67.5-69.5 eV) core-to-valence absorptions are sufficiently separated in photon energy (Figure 4).

Second is the need for a longer time span. In principle, at sufficiently long time delay, the molecules will fully dissociate into isolated atoms, and the atomic ΔA(t) signals will totally converge to a positive plateau. Such convergence is observed for the atomic Br core-to-valence absorptions (Figure 5(c, d)). In contrast, for the atomic core-to-Rydberg absorptions (Figure 10), the signals seem to continue rising at our largest time delay (160 fs). As a result, a longer time span will be necessary to see the full growth of the signals.

Third, the signal-to-noise ratio (S/N)[41,42] of core-to-Rydberg absorptions (Figure 10) is lower than the core-to-valence ones (Figure 5(c, d)), because the XUV spectrum (Figure S1, Supporting Information) is weaker beyond 72 eV due to the cutoff of the Al filter. In this work, one Al filter blocks the driving field before the sample cell and one Zr filter blocks the pump before the grating and XUV camera. It may be more favorable to use two Zr filters to avoid the loss of XUV intensity beyond the Al cutoff (> 72 eV).

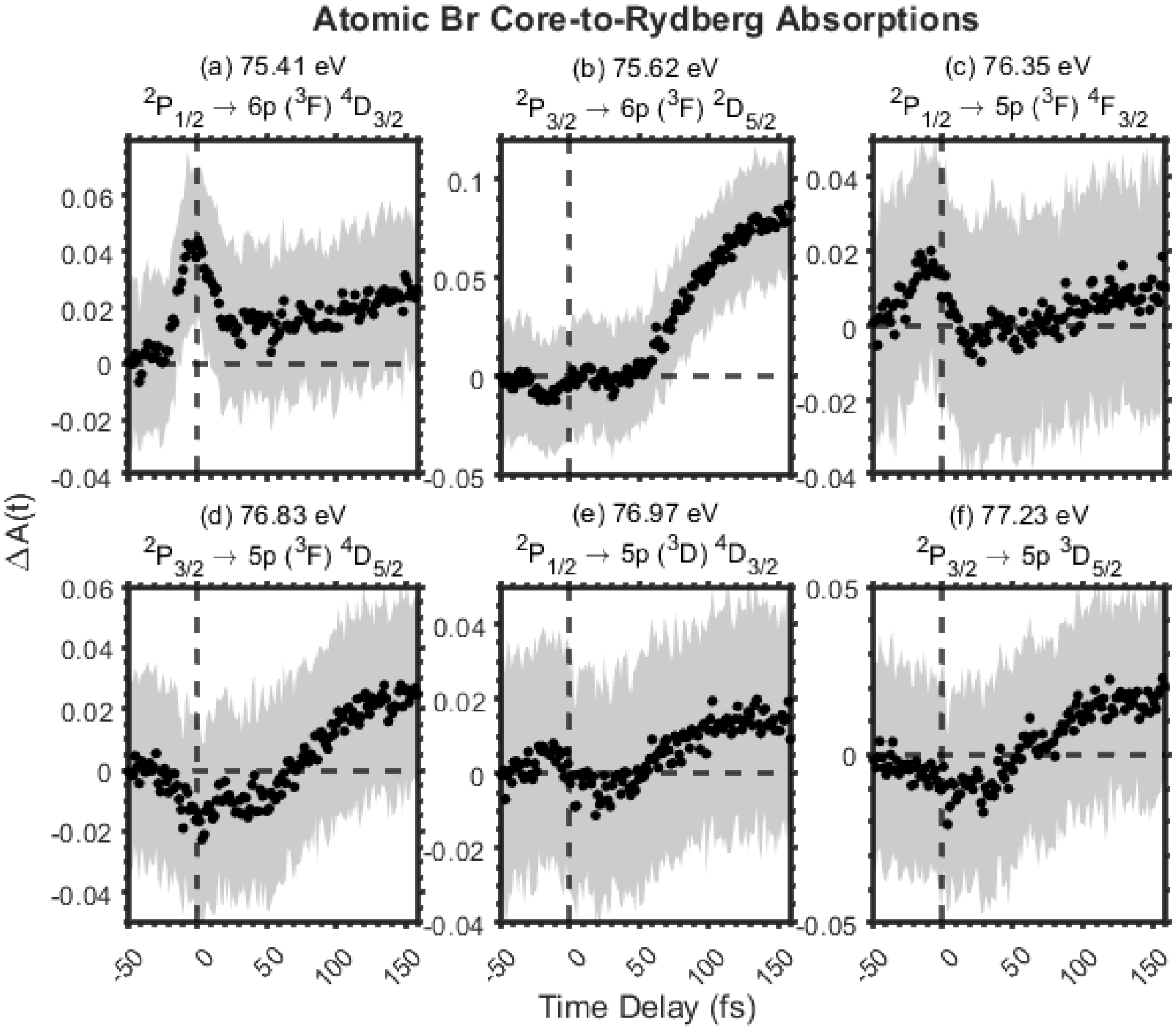

Figure 10. Differential absorption ΔA(t) signal as a function of time delay for selected atomic Br core-to-Rydberg absorptions in Figure 8(c) and Table 6. Black dot and gray shade: best estimate (mean) and uncertainty (one standard deviation)[31] of 93 measurements.

Table 6. Atomic Br core-to-Rydberg 3d → *n*p ($n \geq 5$) absorptions (Figure 8), assigned according to Ref. [20].

| Peak | Atomic Transition[20] | E (eV) (Theory) Ref. [20] | E (eV)[a, b] Ref. [19,20] (Figure 8(d)) | E (eV)[c] Ref. [20] (Figure 8(e)) | E (eV)[d] This work (Figure 8(c)) |
|---|---|---|---|---|---|
| 1 | $^2P_{3/2} \rightarrow 5p\ (^3D)\ ^2D_{5/2}$ | 75.42 | 75.35 | 75.39 | 75.41 |
| 2 | $^2P_{1/2} \rightarrow 5p\ (^3D)\ ^4D_{3/2}$ | 75.63 | 75.60 | 75.56 | 75.62 |
| | $^2P_{3/2} \rightarrow 5p\ (^3F)\ ^2D_{5/2}$ | 75.66 | | | |
| 3 | $^2P_{3/2} \rightarrow 5p\ (^3F)\ ^4D_{5/2}$ | 76.27 | 76.30 | 76.22 | 76.35 |
| | $^2P_{3/2} \rightarrow 5p\ (^3F)\ ^4F_{5/2}$ | 76.35 | | | |
| | $^2P_{3/2} \rightarrow 5p\ (^3P)\ ^2D_{5/2}$ | 76.67 | | | |
| 4 | $^2P_{1/2} \rightarrow 5p\ (^3F)\ ^4F_{3/2}$ | 76.70 | 76.70 | 76.70 | 76.83 |
| | $^2P_{3/2} \rightarrow 6p\ (^3D)\ ^2P_{3/2}$ | 76.72 | | | |
| | $^2P_{3/2} \rightarrow 5p\ (^3P)\ ^2S_{1/2}$ | 76.73 | | | |
| | $^2P_{1/2} \rightarrow 5p\ (^3P)\ ^4D_{3/2}$ | 76.76 | | | |
| 5 | $^2P_{3/2} \rightarrow 6p\ (^3F)\ ^2D_{5/2}$ | 76.93 | 76.85 | 76.85 | 76.97 |
| | $^2P_{3/2} \rightarrow 5p\ (^1F)\ ^2D_{5/2}$ | 77.06 | | | |
| 6 | $^2P_{1/2} \rightarrow 6p\ (^3F)\ ^4D_{3/2}$ | 77.20 | 77.15 | 77.16 | 77.23 |
| 7 | $^2P_{1/2} \rightarrow 7p\ (^3F)\ ^4F_{3/2}$ | 78.49 | 78.40 | | 78.72 |

[a] Measured spectrum in Ref. [19] with numeric values reported in Ref. [20]

[b] Absorption spectroscopy[19] (measurement accuracy ±0.03 eV)

[c] Photoelectron spectroscopy[20] (average photon-energy resolution 0.18 eV)

[d] XUV absorption spectroscopy

## 6. CONCLUSIONS

In this work, the buildup of atomic Br 3d core level absorptions is used to estimate the progress of dissociation. Notably, our measured rise time depends on the probed final state: 38 ± 1 and 20 ± 5 fs for core-to-valence absorptions (Br 3d $^2P_{3/2} \rightarrow$ 4p $^2D_{5/2}$ and $^2D_{3/2}$ at 64.31 and 65.34 eV). On the other hand, the rise times are not determined for core-to-Rydberg absorptions due to spectral overlaps with ground-state bleaching.

With the aid of simulations, our results suggest that the measured probing signals contain not only the wavepacket dynamics but also how each probing transition changes differently with bond length. These potential probe-dependent effects should be considered when interpreting measured signals and their timescales. This may partially explain why the reported "dissociation time" are so different from various experimental probing methods in the literature[2–9] (Table 1).

In addition, this work maps out thoroughly the evolution of the Br 3d core-to-valence and core-to-Rydberg absorptions from molecular to atomic limits in time. Our time-resolved spectrum bridges the static spectra reported separately for $Br_2$ molecules and Br atoms across the scattered literature[15–21] (Table 2).

## ACKNOWLEDGMENT

J.E.B., J.H.O., and S.R.L. acknowledge the financial support from the National Science Foundation under grant CHE-2243756. J.H.O. acknowledges the Government Scholarship to Study Abroad, Ministry of Education, Taiwan for supplemental support, and thanks John Hack and Christian Schroeder for discussions. Y.K. acknowledges startup funding from the University of Michigan. Our calculations used Molecular Graphics and Computation Facility, University of California, Berkeley (funded by National Institutes of Health grant S10OD034382) and we thank Kathleen Durkin, Azhagiya Singam, and Dave Small for support on workstations.

## SUPPORTING INFORMATION

S1. Experimental Method: Time-Resolved XUV Absorption Spectroscopy

S1.1 Broadband XUV Probe Pulse

S1.2 Femtosecond 400 nm Pump Pulse

S1.3 Time-Resolved XUV Absorption Spectroscopy

S2. Computational Method

S2.1 Electronic Structure Calculations

S2.2 Simulation: Nuclear Wavepacket Dynamics

S2.3 Simulation: Time-Resolved XUV Absorption Spectrum

S3. Potential Pathway for Dissociation Product Br ($^2P_{3/2}$) + Br* ($^2P_{1/2}$)

S4. Ground State Bleaching: Coherent Vibrational Dynamics in the Electronic Ground State

S5. The Relative Polarization of 400 nm Pump and XUV Probe Pulses and its Potential Implications

## PRESENT ADDRESS

†John E. Beetar's present address: NSF National eXtreme Ultrafast Science (NeXUS) Facility, Columbus, Ohio 43210, USA

**TOC Graphic**

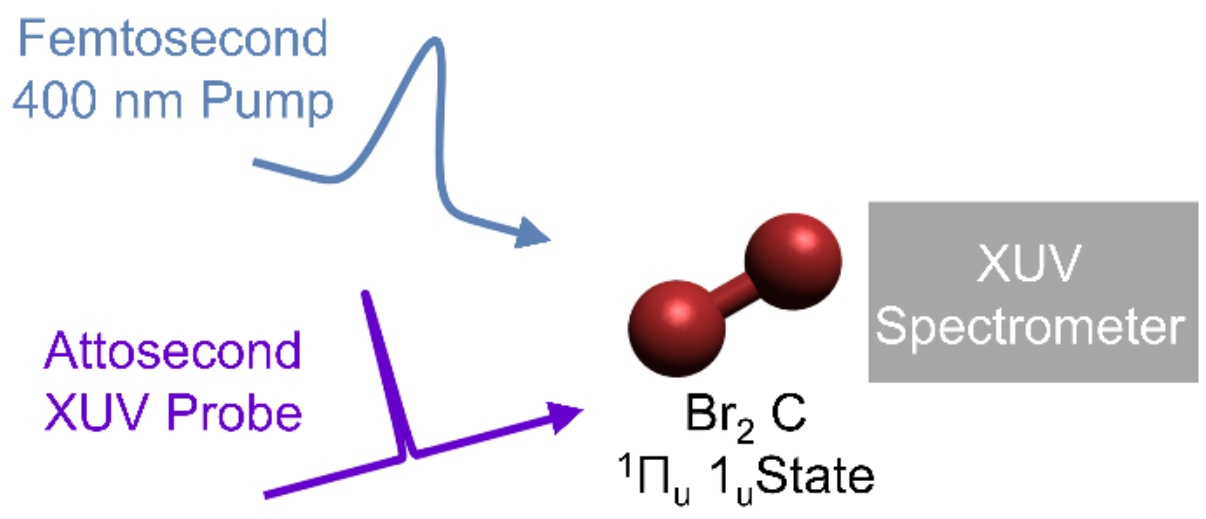